\documentclass[epsfig,12pt]{article}
\usepackage{amssymb}
\usepackage{amsfonts}
\usepackage{graphicx}
\usepackage{amsmath}

\input epsf.sty
\newtheorem{theorem}{Theorem}
\newtheorem{acknowledgement}[theorem]{Acknowledgement}

\input{tcilatex}
\makeatletter \@addtoreset{equation}{section}
\renewcommand{\theequation}{\thesection.\arabic{equation}}

\begin{document}

\title{%
\rightline{\mbox {\normalsize
{Lab/UFR-HEP0512/GNPHE/0514/VACBT/0514}}} \textbf{D-string fluid in conifold:%
}\\
\textbf{I. Topological gauge model}}
\author{R. Ahl Laamara$^{1,2}$, L.B Drissi$^{1,2}$, E.H Saidi$^{1,2,3}$%
\thanks{%
h-saidi@fsr.ac.ma} \\
{\small \textit{1.}} {\small \textit{Lab/UFR-Physique des Hautes Energies,
Facult\'{e} des Sciences de Rabat, Morocco.}}\\
{\small \textit{2. Groupement National de Physique des Hautes Energies,
GNPHE; }}\\
{\small \textit{Siege focal, Lab/UFR-HEP, Rabat, Morocco.}}\\
{\small \textit{3. VACBT, Virtual African Centre for Basic Science and
Technology, }}\\
{\small \textit{Focal point Lab/UFR-PHE, Fac Sciences, Rabat, Morocco.}}}
\maketitle

\begin{abstract}
Motivated by similarities between quantum Hall systems \`{a} la Susskind and
aspects of topological string theory on conifold as well as results obtained
in hep-th/0601020, we study the dynamics of D-string fluids running in
deformed conifold in presence of a strong and constant RR background
B-field. We first introduce the basis of D-string system in fluid
approximation and then derive the holomorphic non commutative gauge
invariant field action describing its dynamics in conifold. This study may
be also viewed as embedding Susskind description for Laughlin liquid in type
IIB string theory. FQH systems on real manifolds $R\times S^{2}$ and $S^{3}$
are shown to be recovered by restricting conifold to its Lagrangian
sub-manifolds. Aspects of quantum behaviour of the string fluid are
discussed. \bigskip \newline
\end{abstract}
\newpage
\tableofcontents
\newpage
\section{Introduction}

\qquad Since Susskind proposal on fractional quantum Hall (FQH) fluids in
Laughlin state as systems described by $\left( 2+1\right) $ non commutative
CS gauge theory $\cite{1}$, there has been a great interest for building new
solutions extending this idea $\cite{2}$-$\cite{6}$. Motivated by: (a)
results concerning attractor mechanism on flux compactification $\cite{7,8}$%
, in particular the link with non commutative geometry, and (b) the study of
$\cite{9}$ dealing with topological non commutative gauge theory on
conifold, we develop in this paper a new extension of Susskind proposal for
FQH fluids to higher dimensions. Our extension deals with modelization of
the dynamics of a fluid of D strings running in conifold and in presence of
a strong and constant RR background B-field. The extended system lives in
complex three (real six) dimensions and is related to the usual FQH system
with point like particles by the following correspondence:

(\textbf{1}) The role of the usual FQH particles moving in a real Riemann
surface $\mathcal{M}$ with coordinates $z$ and $\overline{z}$, is played by
D strings moving on K3 surface with some complex holomorphic coordinates $u$
and $v$ to be specified later. In this picture, FQH particles may be then
viewed as D0 branes coming from D1 strings wrapped on $S^{1}$.

(\textbf{2}) The complex coordinates $z_{a}\left( t\right) $ and $\overline{z%
}_{a}\left( t\right) $ parameterizing the dynamics of the N fractional
quantum Hall particles are then mapped to $u_{a}=u_{a}\left( \xi \right) $
and $v_{a}=v_{a}\left( \xi \right) $ with $\xi =t+i\sigma $ being the string
world sheet complex coordinate.

(\textbf{3}) The local coordinates $\left( t,z,\overline{z}\right) $
parameterize a real three dimension space; say the space $R^{1,2}$. The
local variables $\left( \xi ,u,v\right) $ parameterize a complex three
dimension space, which is just the conifold $T^{\ast }S^{3}$ realized as $%
T^{\ast }S^{1}$ fibered on $T^{\ast }S^{2}$. The $R^{1,2}$ geometry used in
Susskind description appears then as a special real three dimension slice of
conifold.

(\textbf{4}) The role of the magnetic field B is now played by a constant
and strong RR background field $\mathrm{B}$ of type IIB string. Like in FQH
system, the B field is supposed normal to K3 surface and strong enough so
that one can neglect other possible interactions.

\qquad From this naive and rapid presentation of the higher dimensional
extended FQH system, to which we refer here below as a D-string fluid (DSF
for short), one notes some specific properties among which the three
following: First, Susskind proposal may be recovered from DSF by taking
appropriate parameter limits of DSF moduli space to be described later.
Second, the real geometry of FQH system is contained in conifold; the
present study may be then thought of as embedding Susskind field theoretical
model for Laughlin state with filling factor $\nu =\frac{1}{k}$ into type
IIB superstring theory on conifold. This property offers one more argument
for embedding FQH systems in supersymmetric theories; others arguments have
been discussed in $\cite{10,11}$. Finally, in DSF model, the complex
holomorphy property plays a basic role; reality is recovered by restricting
conifold to its half dimension Lagrangian sub-manifold. This involution has
the effect of projecting DSF into the usual FQH system opening the way for
links between real 3D physics and type II superstrings on Calabi-Yau
threefolds.

\qquad The presentation of this paper is as follows: In section 2, we
introduce the basis of fluid approximation of D-strings running in conifold.
To build this system, we use special properties of K3 complex surface and
conifold geometry. We also take advantage of Susskind model for Laughlin
liquid which we use as a reference to make comparisons and physical
interpretations. In section 3, we study the classical dynamics of the
interaction between D strings and the RR magnetic background field. We
suppose that B is strong enough so that one can neglect string kinetic
energy and mutual energy interactions between the D strings. We also suppose
that the number of D strings per volume unit is high and uniform. Then use
the fluid approximation to derive the effective field theory extending
Susskind model. In this section, we also study some special limits such as
real projection. In section 4, we discuss quantum aspects of the D-strings
fluid, in particular holomorphic property and in section 5 we give our
conclusion and outlook.

\section{D fluid Model proposal}

\qquad Like in usual fractional quantum Hall fluids in real three
dimensions, the D string system we consider here involves, amongst others,
two basic ingredients:

(\textbf{a}) A set of N D strings running in conifold and printing a line
trajectory $\mathcal{T}$ on the complex two surface K3. The curve $\mathcal{T%
}$ is exactly the world line trajectory one gets if the D strings $\left(
u\left( \xi \right) ,v\left( \xi \right) \right) $ collapse to point like
particles $\left( z\left( t\right) ,\overline{z}\left( t\right) \right) $.

(\textbf{b}) A constant RR background field \textrm{B}, which is taken
normal to K3, governs the dynamics of the strings. The magnitude of the
\textrm{B} field is supposed strong enough such that one can neglect all
other interactions in the same spirit as we do in FQH systems involving
point like particles. Non zero \textrm{B }field induces then a non
commutative geometry on K3 captured by the Poisson bracket $\left\{ \mathcal{%
X}\left( \xi ,u,v\right) ,\mathcal{Y}\left( \xi ,u,v\right) \right\}
_{u,v}\sim \partial _{u}\mathcal{X}\partial _{v}\mathcal{Y}-\partial _{v}%
\mathcal{X}\partial _{u}\mathcal{Y}$ of the dynamical variables $\mathcal{X}%
\left( \xi ,u,v\right) $ and $\mathcal{Y}\left( \xi ,u,v\right) $ of the
fluid approximation.

\qquad To get the gauge invariant effective field action $\mathcal{S}_{DSF}$
describing the dynamics of fluids of D strings in conifold with analogous
conditions as in FQH systems, we need two essential things. First fix the
classical field variables $u=u\left( \xi \right) $ and $v=v\left( \xi
\right) $ describing the D string dynamics in conifold and second implement
the fluid approximation by using a uniform particle density $\rho =\rho
\left( u,v\right) $\ to deal with the number of D strings per volume unit.
We know how this is done in the case of standard FQH fluids in Laughlin
state with filling fraction $\nu =\frac{1}{k}$ and we would like to extend
this construction for D string fluids taken in similar conditions. Though
the geometries involved in the present study are a little bit complicated
and the basic objects are one dimensional extended elements, we will show
that the theoretical analysis is quite straightforward.

\qquad For the choice of the string variables $u=u\left( \xi \right) $ and $%
v=v\left( \xi \right) $; they are given by the geometry of K3 and the fluid
description is obtained by extending Susskind analysis for FQH particles.
For fluid approximation, we use also properties of holomorphic area
preserving diffeomorphisms on K3 demanding a uniform density. Seen that the
idea of the general picture has been exposed before and seen that details
requires involved tools, we begin the present analysis by describing, in
next subsection, the D string dynamical variables. Then come back to the
fluid approximation with uniform density. More details on the holomorphic
gauge invariant field action and real truncating will be considered in the
forthcoming section.

\subsection{D string variables}

\qquad First note that there are various kinds of K3 surfaces; the one we
will be using below is a local K3 with a deformed $A_{1}$ singularity; that
is $T^{\ast }P^{1}\simeq T^{\ast }S^{2}$. Second note also that the complex
surface K3 is a non flat Kahler manifold and so the natural way to define it
is in term of a projective surface embedded in a homogeneous complex three
space as given below,%
\begin{equation}
xy-zw=\mu ,\qquad x,y,z,w\in \mathbb{C},  \label{1}
\end{equation}%
together with the following projective transformations,%
\begin{equation}
\left( x,y,z,w\right) \qquad \rightarrow \qquad \left( \lambda x,\frac{1}{%
\lambda }y,\lambda z,\frac{1}{\lambda }w\right) ,  \label{2}
\end{equation}%
and where $\mu $ is a complex constant. In these relations, we have four
complex holomorphic variables namely $x,y,z$ and $w$; but not all of them
are free. They are subject to two constraint relations (\ref{1}-\ref{2})
reducing the degrees of freedom down to two. Note in passing that by setting
$y=\overline{x}$ and $w=\overline{z}$, the above relations reduce to%
\begin{eqnarray}
\left\vert x\right\vert ^{2}+\left\vert z\right\vert ^{2} &=&\func{Re}\mu ,
\notag \\
\left( x,z\right) \qquad &\rightarrow &\qquad e^{i\theta }\left( x,z\right) ,
\label{02}
\end{eqnarray}%
so they define a real two sphere $S^{2}$ embedded in complex space $\mathbb{C%
}^{2}$ parameterized by $\left( x,z\right) $. This is an interesting
property valid not only for $T^{\ast }S^{2}$; but also for conifold $T^{\ast
}S^{3}$. This crucial property will be used to recover the hermitian models
on real three dimension space; it deals with the derivation of Lagrangian
sub-manifold from mother manifold $T^{\ast }S^{3}$. As we will see it
progressively, this feature is present everywhere along all of this paper.
We will then keep it in mind and figure it out only when needed to make
comments.

\qquad To implement string dynamics, we should add time variable $t$ and the
string variable $\sigma $ parameterizing the one dimensional D string
geometry. If we were dealing with a point like particle moving on this
complex surface, the variables would be given by the $1d$ fields,%
\begin{equation}
x=x\left( t\right) ,\qquad y=y\left( t\right) ,\qquad z=z\left( t\right)
,\qquad w=w\left( t\right) .  \label{3}
\end{equation}%
For the case of a D string with world sheet variable $\xi =t+i\sigma $
moving on $T^{\ast }P^{1}$, the D string variables are then given by the $2d$
fields,%
\begin{equation}
x=x\left( \xi \right) ,\qquad y=y\left( \xi \right) ,\qquad z=z\left( \xi
\right) ,\qquad w=w\left( \xi \right) .  \label{4}
\end{equation}%
with $\left\vert \sigma \right\vert \leq l$ and obviously the constraint eqs(%
\ref{1}-\ref{2}). In the limit $l\rightarrow 0$, the above $2d$ fields
reduces to the previous one dimensional variables. Since K3 surface as
considered here is a projective algebraic surface using complex holomorphic
variables, it is natural to make the two following hypothesis:

(\textbf{i}) \textbf{Field Holomorphy}: \qquad We suppose that the above
D-string field variables eqs(\ref{4}) have no $\overline{\xi }$ dependence;
that is holomorphic functions in $\xi $,
\begin{equation}
\frac{\partial \phi }{\partial \overline{\xi }}=0,\qquad \phi \left( \xi
\right) =\sum_{n\in Z}\alpha _{n}^{\phi }\xi ^{n},\qquad \phi =x,y,z,w,
\label{5}
\end{equation}%
where $\alpha _{n}^{\phi }=\frac{1}{2\pi i}\doint \frac{d\xi }{\xi ^{n+1}}%
\phi \left( \xi \right) $ are string modes. This hypothesis means that the D
string we are dealing with is either a one handed mover closed D-string, say
a left mover closed string, or an open D-string with free ends. To fix the
ideas, we consider here below closed D-strings and think about $\xi =\exp
\left( \tau +i\widetilde{\sigma }\right) $ with $0\leq \sigma =l\widetilde{%
\sigma }\leq 2\pi l$. Holomorphy hypothesis selects one sector; it requires
that the variables parameterizing the D-strings are complex holomorphic and
same for the field action $\mathcal{S}_{DSF}=\mathcal{S}_{DSF}\left[ x,y,z,w%
\right] $ that describe their dynamics. Usual hermiticity is recovered by
restricting conifold to its Lagrangian sub-manifold obtained by setting $\xi
=\overline{\xi },$ $y=\overline{x}$ and $w=\overline{z}$.

(\textbf{ii}) \textbf{Induced gauge symmetry}:\qquad For later use it is
interesting to treat on equal footing the string world sheet variable $\xi $
and those parameterizing K3. This may be done by thinking about the
projective transformations (\ref{2}) also as those one gets by performing
the change,%
\begin{equation}
\xi \qquad \rightarrow \qquad \lambda \xi ,  \label{6}
\end{equation}%
with $\lambda $\ a non zero complex parameter. In other words, the string
variables obey the following,%
\begin{eqnarray}
x\left( \lambda \xi \right) &=&\lambda x\left( \xi \right) ,\qquad z\left(
\lambda \xi \right) =\lambda z\left( \xi \right) ,  \notag \\
y\left( \lambda \xi \right) &=&\frac{1}{\lambda }y\left( \xi \right) ,\qquad
w\left( \lambda \xi \right) =\frac{1}{\lambda }w\left( \xi \right) ,
\label{7}
\end{eqnarray}%
together with the local constraint eqs,%
\begin{equation}
x\left( \xi \right) y\left( \xi \right) -z\left( \xi \right) w\left( \xi
\right) =\mu .  \label{8}
\end{equation}%
Note that eq(\ref{8}) describes in fact an infinite set of constraint
relations since for each value of $\xi \in \mathbb{C}^{\ast }$, the D-string
fields should obey (\ref{8}). This feature has a nice geometric
interpretation. The string dynamics involves five complex holomorphic
variables namely $\left( \xi ,x,y,z,w\right) $ and the two algebraic
constraint equations (\ref{1}-\ref{2}). Therefore these variables
parameterize a complex three dimension projective hypersurface embedded in $%
\mathbb{C}^{5}$ and which is nothing else that the deformed conifold
geometry $T^{\ast }S^{3}$ with the realization,%
\begin{equation}
T^{\ast }S^{3}\simeq T^{\ast }S^{1}\times T^{\ast }S^{2}.  \label{9}
\end{equation}%
In this fibration, $T^{\ast }S^{2}$ is the base sub-manifold and the fiber $%
T^{\ast }S^{1}$ describes the D-string world sheet.

\qquad To summarize, the variables describing the motion of a D-string in
conifold are given by eqs(\ref{6}-\ref{8}). For a system of N D-strings
moving in conifold, we have then,%
\begin{equation}
x_{a}\left( \xi \right) y_{a}\left( \xi \right) -z_{a}\left( \xi \right)
w_{a}\left( \xi \right) =\mu ,\qquad a=1,...,N,  \label{10}
\end{equation}%
where for each value of the index a, we have also the eqs (\ref{6}-\ref{7}).
Having fixed the variables, we turn now to describe the fluid approximation
of D-strings and implement the constant and strong background RR B-field.

\subsection{Fluid approximation}

\qquad For later analysis, it is convenient to use the usual $SL\left(
2\right) $ isometry of the conifold to put the above relations into a
condensed form. Setting
\begin{equation}
X^{i}=\left( x\left( \xi \right) ,z\left( \xi \right) \right) ,\qquad
Y_{i}=\left( y\left( \xi \right) ,w\left( \xi \right) \right) ,  \label{11}
\end{equation}%
transforming as isodoublets under $SL\left( 2\right) $\ isometry, the
coordinates of a given D string moving in conifold is given by the
holomorphic field doublets,%
\begin{equation}
X^{i}=X^{i}\left( \xi \right) ,\qquad Y_{i}=Y_{i}\left( \xi \right) ,\qquad
i=1,2,  \label{12}
\end{equation}%
with the local constraint eqs,
\begin{equation}
\epsilon _{ij}X^{i}\left( \xi \right) Y^{j}\left( \xi \right) =\mu ,
\label{13}
\end{equation}%
and the projective symmetry
\begin{eqnarray}
X^{i}\left( \lambda \xi \right) &=&\lambda X^{i}\left( \xi \right) ,  \notag
\\
Y_{i}\left( \lambda \xi \right) &=&\frac{1}{\lambda }Y_{i}\left( \xi \right)
.  \label{14}
\end{eqnarray}%
Using these notations, the system of D string reads then as follows%
\begin{equation}
\epsilon _{ij}X_{a}^{i}\left( \xi \right) Y_{a}^{j}\left( \xi \right) =\mu
,\qquad a=1,...,N,  \label{15}
\end{equation}%
where $\epsilon _{ij}$ is the usual two dimensional antisymmetric tensor
with $\epsilon _{12}=1$. In the large $N$ limit with density $\rho \left(
\xi ,x,y\right) $; i.e $N=\int_{T^{\ast }S^{3}}d^{3}v\rho \left( \xi
,x,y\right) ,$ where $\left( x,y\right) $ sometimes denoted also as $\left(
x^{1},x^{2},y_{1},y_{2}\right) $ stand for the pairs of doublets $\left(
X^{i},Y^{j}\right) $, the D-string system may be thought of as a fluid of D1
branes running in conifold. Along with the previous relations, the fluid
approximation allows the following substitutions,%
\begin{eqnarray}
\left\{ X_{a}^{i}\left( \xi \right) ,\text{ }1\leq a\leq N\right\} \qquad
&\rightarrow &\qquad \mathcal{X}^{i}\left( \xi ,x,y\right) ,  \notag \\
\left\{ Y_{a}^{i}\left( \xi \right) ,\text{ }1\leq a\leq N\right\} \qquad
&\rightarrow &\qquad \mathcal{Y}^{i}\left( \xi ,x,y\right) ,  \label{16}
\end{eqnarray}%
together with eqs(\ref{15}) replaced by
\begin{equation}
\epsilon _{ij}\mathcal{X}^{i}\mathcal{Y}^{j}=\mu ,\qquad \mathcal{X}^{i}=%
\mathcal{X}^{i}\left( \xi ,x,y\right) ,\qquad \mathcal{Y}^{i}=\mathcal{Y}%
^{i}\left( \xi ,x,y\right) ,  \label{17}
\end{equation}%
and projective symmetry promoted to,%
\begin{eqnarray}
\mathcal{X}^{i}\left( \lambda \xi ,\lambda x,\frac{1}{\lambda }y\right)
&=&\lambda \mathcal{X}^{i}\left( \xi ,x,y\right) ,  \notag \\
\mathcal{Y}^{i}\left( \lambda \xi ,\lambda x,\frac{1}{\lambda }y\right) &=&%
\frac{1}{\lambda }\mathcal{Y}^{i}\left( \xi ,x,y\right) .  \label{18}
\end{eqnarray}%
For physical interpretation, we will also use the splitting%
\begin{equation}
\mathcal{X}^{i}=x^{i}+\mu \mathcal{C}_{+}^{i},\qquad \mathcal{Y}%
_{i}=y_{i}-\mu \mathcal{C}_{-i},
\end{equation}%
where $\mathcal{C}_{+}^{i}$ and $\mathcal{C}_{-i}$ are gauge fields
constrained as
\begin{equation}
x^{i}\mathcal{C}_{-i}-y_{i}\mathcal{C}_{+}^{i}+\mu \mathcal{C}_{-i}\mathcal{C%
}_{+}^{i}=0,  \label{b}
\end{equation}%
scaling as the inverse of length and describing fluctuations around the
static positions $x^{i}$ and $y_{i}$. From $SL\left( 2\right) $
representation theory, one may also split the fields $\mathcal{X}^{i}$ and $%
\mathcal{Y}^{i}$ using holomorphic vielbein gauge fields,
\begin{eqnarray}
\mathcal{X}^{i}\left( \xi ,x,y\right) &=&x^{i}E_{+-}+\epsilon
^{ij}y_{j}A_{++},  \notag \\
\mathcal{Y}_{i}\left( \xi ,x,y\right) &=&y_{i}E_{-+}-\epsilon
_{ij}x^{j}A_{--},  \label{19}
\end{eqnarray}%
where $E_{\pm \mp }$ should be as $E_{\pm \mp }=\left( 1+A_{\pm \mp }\right)
$. Like for $\mathcal{X}^{i}$ and $\mathcal{Y}^{i}$, the gauge fields $%
\mathcal{C}_{+}^{i}$ and $\mathcal{C}_{-i}$ as well as $E_{\pm \mp }$ and $%
A_{\pm \pm }$ are homogeneous holomorphic functions subject to the
projective transformations $\mathcal{C}_{\pm }\left( \lambda \xi ,\lambda x,%
\frac{1}{\lambda }y\right) =\lambda ^{\pm }\mathcal{C}_{\pm }\left( \xi
,x,y\right) $ and,
\begin{eqnarray}
E_{+-}\left( \lambda \xi ,\lambda x,\frac{1}{\lambda }y\right)
&=&E_{+-}\left( \xi ,x,y\right) ,  \notag \\
A_{++}\left( \lambda \xi ,\lambda x,\frac{1}{\lambda }y\right) &=&\lambda
^{2}A_{++}\left( \xi ,x,y\right) ,  \notag \\
E_{-+}\left( \lambda \xi ,\lambda x,\frac{1}{\lambda }y\right)
&=&E_{-+}\left( \xi ,x,y\right) ,  \label{pr} \\
A_{--}\left( \lambda \xi ,\lambda x,\frac{1}{\lambda }y\right) &=&\lambda
^{-2}A_{--}\left( \xi ,x,y\right) .  \notag
\end{eqnarray}%
Using the conifold defining relation $\epsilon _{ij}\mathcal{X}^{i}\mathcal{Y%
}^{j}=\mu $, we see that, like for $\mathcal{C}_{\pm }^{i}$ gauge fields,
the above holomorphic vielbeins capture two complex degrees of freedom only
since in addition to eqs(\ref{pr}), they satisfy moreover,
\begin{equation}
E_{+-}E_{-+}-A_{++}A_{--}=1.  \label{c}
\end{equation}%
An equivalent relation using $A_{\pm \mp }$ and $A_{\pm \pm }$ may be also
written down. As far as the constraint eq(\ref{b},\ref{c}) are concerned,
there are more than one way to deal with. One way is to solve it
perturbatively as $E_{\pm \mp }\simeq \left( 1+A_{\pm \mp }\right) $ with $%
A_{\pm \mp }=\pm A_{0}$ and then substitute $A_{0}=i\sqrt{A_{++}A_{--}}$. An
other way is to solve eq(\ref{c}) exactly as
\begin{eqnarray}
E_{+-} &=&K\sqrt{1+A_{++}A_{--}},  \notag \\
E_{-+} &=&\frac{1}{K}\sqrt{1+A_{++}A_{--}},
\end{eqnarray}%
where $K$ is an arbitrary non zero function. In both cases one looses field
linearity which we would like to have it. We will then keep the gauge field
constraint eqs as they are and give the results involving all these
components using Lagrange method. Notice that, from physical view, the gauge
fields $\mathcal{C}_{\pm }^{i}$ or equivalently $A_{\pm \mp }=A_{\pm \mp
}\left( \xi ,x,y\right) $ and $A_{\pm \pm }=A_{\pm \pm }\left( \xi
,x,y\right) $ describe gauge fluctuations around the static solution
\begin{equation}
\mathcal{X}^{i}=x^{i},\qquad \mathcal{Y}_{i}=y_{i},\qquad x^{i}y_{i}=\mu ,
\label{20}
\end{equation}%
preserving conifold volume 3-form. Expressing the field $\mathcal{X}^{i}$
and $\mathcal{Y}_{i}$ as $\mathcal{X}^{i}=x^{i}+\mu \mathcal{C}_{+}^{i}$ and
$\mathcal{Y}_{i}=y_{i}-\mu \mathcal{C}_{-i}$, we have $\mu \mathcal{C}%
_{+}^{i}=x^{i}A_{+-}+y^{i}A_{++}$ and $\mu \mathcal{C}%
_{-i}=-y_{i}A_{-+}+x_{i}A_{--}$. Notice also that, as general coordinate
transformations, the splitting (\ref{19}) may be also defined as holomorphic
diffeomorphisms $\mathcal{X}^{i}=\mathcal{L}_{v}x^{i}$ and $\mathcal{Y}_{i}=%
\mathcal{L}_{v}y_{i}$ where the vector field $\mathcal{L}_{v}$ is given by
\begin{equation}
\mathcal{L}_{v}=V_{++}D_{--}+V_{--}D_{++}+V_{0}D_{0}+V_{0}^{\prime }\Delta
_{0},  \label{21}
\end{equation}%
with gauge component fields $V_{pq},$ $p,$ $q=+,-$ and where the
dimensionless derivatives generating the $GL\left( 2\right) $ group are
given by,%
\begin{equation}
\Delta _{0}=\frac{1}{2}\left( x^{i}\frac{\partial }{\partial x^{i}}+y^{i}%
\frac{\partial }{\partial y^{i}}\right) ,
\end{equation}%
or naively\ as $\Delta _{0}=\frac{\partial }{\partial \left(
x^{i}y_{i}\right) }$, and
\begin{equation}
D_{--}=y^{i}\frac{\partial }{\partial x^{i}},\qquad D_{++}=x^{i}\frac{%
\partial }{\partial y^{i}},\qquad D_{0}=\left( x^{i}\frac{\partial }{%
\partial x^{i}}-y^{i}\frac{\partial }{\partial y^{i}}\right) .  \label{38}
\end{equation}%
In these eqs, we have two charge operators; the operator $\Delta _{0}$
generates the abelian scaling factor with the property%
\begin{equation}
\left[ \Delta _{0},D_{\pm \pm }\right] =2D_{\pm \pm },\qquad \left[ \Delta
_{0},D_{0}\right] =0,
\end{equation}%
and $D_{0}=\left[ D_{++},D_{--}\right] $ generates the abelian Cartan Weyl $%
GL\left( 1\right) $ subgroup of $SL\left( 2\right) $. Notice moreover that
inverting the decomposition (\ref{19}), we can write the vielbein fields as
follows,%
\begin{eqnarray}
E_{+-} &=&\frac{1}{\mu }y_{i}\mathcal{X}^{i}=1+y_{i}\mathcal{C}%
_{+}^{i},\qquad E_{-+}=\frac{1}{\mu }\mathcal{Y}_{i}x^{i}=1-x^{i}\mathcal{C}%
_{-i},  \notag \\
A_{++} &=&\frac{1}{\mu }\epsilon _{ij}x^{i}\mathcal{X}^{j}=-x_{i}\mathcal{C}%
_{+}^{i},\qquad A_{--}=\frac{1}{\mu }\epsilon ^{ij}y_{i}\mathcal{Y}%
_{j}=-y^{i}\mathcal{C}_{-i},  \label{23}
\end{eqnarray}%
As one sees, these gauge fluctuations $E_{\pm \mp }$ and $A_{\pm \pm }$ are
dimensionless; they let understand that they should appear as gauge fields
covariantizing dimensionless linear differential operators. These are just
the $D_{0,\pm \pm }$ operators given above. At the static point eq(\ref{20}%
), we also see that $E_{+-}=E_{-+}=1$ and $A_{\pm \pm }=0$, ($\mathcal{C}%
_{\pm }^{i}=0$). With these tools we are now in position to address the
building of the effective field action of the D string fluid model in
conifold.

\section{Field action}

\qquad To get the gauge invariant effective field action $\mathcal{S}_{DSF}=%
\mathcal{S}_{DSF}\left[ \mathcal{C}_{\pm }^{i},\mathcal{C}_{0}\right] $
describing the dynamics of the D string fluid in the conifold, we borrow
ideas from Susskind method used for FQH liquid of point like particles. We
first give the classical field action $\mathcal{S}_{clas}\left[ X,Y\right] $
describing the interaction between a given D string $\left\{ X\left( \xi
\right) ,Y\left( \xi \right) \right\} $ moving in the RR background field B.
Then we consider the fluid approximation using the field variables $\left\{
\mathcal{X}\left( \xi ,x,y\right) ,\mathcal{Y}\left( \xi ,x,y\right)
\right\} $ instead of the coordinates $\left\{ X_{a}\left( \xi \right)
,Y_{a}\left( \xi \right) ,\text{ \ }1\leq a\leq N\right\} $. In this limit
we suppose that density $\rho \left( \xi ,x,y\right) $ is large and uniform;
i.e $\rho \left( \xi ,x,y\right) =\rho _{0}$. Finally, we derive the
effective gauge field action once by using the D-string field variables $%
\mathcal{X}$ and $\mathcal{Y}$; i.e $\mathcal{S}=\mathcal{S}_{DSF}\left[
\mathcal{X},\mathcal{Y},\mathcal{\Lambda }\right] $ and an other time by
using gauge fields $\mathcal{C}_{\pm }^{i}$ and $\mathcal{C}_{0}$ describing
the fluctuations around the static positions.

\subsection{Classical B$_{RR}$-D string coupling}

\qquad To start recall that the field action $\mathcal{S}_{clas}\left[ z,%
\overline{z}\right] =\int dt\mathcal{L}_{clas}\left( z,\overline{z}\right) $
describing the classical dynamics of a charged particle with coordinate
positions $z=z\left( t\right) $ and $\overline{z}=\overline{z}\left(
t\right) $, in a constant and strong background magnetic field B, is given
by
\begin{equation}
\mathcal{L}_{clas}=\frac{iB}{2}\left( \overline{z}\left( t\right) \frac{%
dz\left( t\right) }{dt}-z\left( t\right) \frac{d\overline{z}\left( t\right)
}{dt}\right) .
\end{equation}%
For a system of N classical D-strings $\left\{ X_{a}\left( \xi \right)
,Y_{a}\left( \xi \right) ,\text{ \ }1\leq a\leq N\right\} $ in the RR
background magnetic field, one has a quite similar quantity. The above point
like particle action extends as follows,%
\begin{equation}
\mathcal{S}_{N}\left[ X,Y\right] =\frac{1}{2}\int_{T^{\ast }S^{1}}d\xi
\sum_{a=1}^{N}B_{ij}\left( Y_{a}^{j}\left( \xi \right) \frac{\partial
X_{a}^{i}\left( \xi \right) }{\partial \xi }-Y_{a}^{i}\left( \xi \right)
\frac{\partial X_{a}^{j}\left( \xi \right) }{\partial \xi }\right) ,
\label{24}
\end{equation}%
with $B_{ij}=i\mathrm{B}\epsilon _{ij}$ and which, for convenience, we
rewrite also as%
\begin{equation}
\mathcal{S}_{N}\left[ X,Y\right] =\frac{i\mathrm{B}}{2}\int_{T^{\ast
}S^{1}}d\xi \sum_{a=1}^{N}\left( Y_{ia}\frac{\partial X_{a}^{i}}{\partial
\xi }-X_{a}^{i}\frac{\partial Y_{ia}}{\partial \xi }\right) .  \label{25}
\end{equation}%
This field action $\mathcal{S}_{N}\left[ X,Y\right] $ exhibits three special
and remarkable features; first it is holomorphic and the corresponding
hermitian $\mathcal{S}_{N}^{\func{real}}\left[ X,\overline{X}\right] $
follows by setting,
\begin{equation}
Y_{ia}=\overline{\left( X_{a}^{i}\right) },\qquad \xi =\overline{\xi }%
=t,\qquad \mathrm{B}=\overline{\mathrm{B}}.  \label{26}
\end{equation}%
As such we have%
\begin{equation}
\mathcal{S}_{N}^{\func{real}}\left[ X,\overline{X}\right] =\frac{i\func{Re}%
\mathrm{B}}{2}\int dt\sum_{a=1}^{N}\left( \overline{\left( X_{a}^{i}\right) }%
\frac{dX_{a}^{i}}{dt}-X_{a}^{i}\frac{d\overline{\left( X_{a}^{i}\right) }}{dt%
}\right) .  \label{27}
\end{equation}%
The second feature of $\mathcal{S}_{N}\left[ X,Y\right] $ deals with the
hypersurface eq(\ref{15}). Since $Y_{ia}X_{a}^{i}=\mu $ is a constraint eq
on the dynamical field variables, it can be implemented in the action by
using a Lagrange gauge field $\Lambda =\Lambda \left( \xi \right) $. So eq(%
\ref{24}) should be read as,
\begin{eqnarray}
\mathcal{S}_{N}\left[ X,Y,\Lambda \right] &=&\frac{i\mathrm{B}}{2}%
\int_{T^{\ast }S^{1}}d\xi \sum_{a=1}^{N}\left( Y_{ia}\frac{\partial X_{a}^{i}%
}{\partial \xi }-X_{a}^{i}\frac{\partial Y_{ia}}{\partial \xi }\right)
\notag \\
&&+\frac{\mathrm{B}}{2}\int_{T^{\ast }S^{1}}d\xi \sum_{a=1}^{N}\Lambda
_{a}\left( \xi \right) \left( Y_{ia}\left( \xi \right) X_{a}^{i}\left( \xi
\right) -\mu \right) .  \label{28}
\end{eqnarray}%
The difference between $\mathcal{S}_{N}\left[ X,Y\right] $ of eq(\ref{25})
and the above $\mathcal{S}_{N}\left[ X,Y,\Lambda \right] $ is that in the
second description the field variables $X_{a}^{i}\left( \xi \right) $ and $%
Y_{a}^{j}\left( \xi \right) $ are unconstrained. Conifold target
hypersurface is obtained by minimizing $\mathcal{S}_{N}\left[ X,Y,\Lambda %
\right] $ with respect to $\Lambda ,$
\begin{equation}
\frac{\delta S_{N}\left[ X,Y,\Lambda \right] }{\delta \Lambda _{a}}%
=Y_{ia}\left( \xi \right) X_{a}^{i}\left( \xi \right) -\mu =0.
\end{equation}%
The third feature concerns the computation of the conjugate momentum $\Pi
_{i}=\frac{\partial \mathcal{L}}{\partial \left( \partial X^{i}/\partial \xi
\right) }$ of the field variable $X^{i}$. One discovers that the coordinate
variables $Y_{i}$ and $X^{i}$ are conjugate fields. This property shows that
the underlying conifold geometry with the background field behaves as a non
commutative manifold.

\qquad Notice that, as required by the construction, eq(\ref{24}) is
invariant under the global symmetry
\begin{eqnarray}
\xi \qquad &\rightarrow &\qquad \lambda \xi ,  \notag \\
X_{a}^{i}\qquad &\rightarrow &\qquad \lambda X_{a}^{i},  \label{29} \\
Y_{a}^{i}\qquad &\rightarrow &\qquad \frac{1}{\lambda }Y_{a}^{i},  \notag
\end{eqnarray}%
with $\frac{d\lambda }{d\xi }=0$. This is a crucial point as far as we are
thinking about conifold as given by the fibration $T^{\ast }S^{1}\times
T^{\ast }S^{2}$. Now, using the fluid approximation mapping the system $%
\left\{ X_{a}^{i}\left( \xi \right) ,\text{ }Y_{a}^{j}\left( \xi \right) ,%
\text{ }\Lambda _{a}\left( \xi \right) \text{ };1\leq a\leq N\right\} $ into
the 3D holomorphic fields $\mathcal{X}^{i}=\mathcal{X}^{i}\left( \xi
,x,y\right) ,$ $\mathcal{Y}^{j}=\mathcal{Y}^{j}\left( \xi ,x,y\right) $ and $%
\mathcal{\Lambda }=\mathcal{\Lambda }\left( \xi ,x,y\right) $, we can put eq(%
\ref{28}) as a complex 3D holomorphic field action
\begin{equation}
\mathcal{S}_{2}\left[ \mathcal{X},\mathcal{Y},\mathcal{\Lambda }\right]
=\int_{T^{\ast }S^{3}}d^{3}v\mathcal{L}_{2}\left( \mathcal{X},\mathcal{Y},%
\mathcal{\Lambda }\right) ,
\end{equation}%
with%
\begin{equation}
\mathcal{L}_{2}\left( \mathcal{X},\mathcal{Y},\mathcal{\Lambda }\right) =%
\frac{i\mathrm{B}}{2\mu }\left[ \left( \mathcal{Y}_{i}\partial _{0}\mathcal{X%
}^{i}-\mathcal{X}^{i}\partial _{0}\mathcal{Y}_{i}\right) -i\Lambda \left(
\mathcal{Y}_{i}\mathcal{X}^{i}-\mu \right) \right] ,  \label{30}
\end{equation}%
and $\partial _{0}=\xi \frac{\partial }{\partial \xi }=\frac{\partial }{%
\partial \ln \xi }$ and where $d^{3}v$ is the conifold holomorphic volume
measure given by,%
\begin{equation}
d^{3}v=\frac{d\xi \wedge dx^{i}\wedge dy_{i}}{\xi },\qquad x^{i}y_{i}=\mu .
\label{vol}
\end{equation}%
For more details on the specific properties of this complex volume see $\cite%
{12}$; for the moment let us push forward this description using the $%
T^{\ast }S^{1}\times T^{\ast }S^{2}$ realization of conifold. In this view,
notice that on $T^{\ast }S^{1}$, the global holomorphic operator $\partial
=d\xi \frac{\partial }{\partial \xi }$ may be also written as $\partial
=d\varsigma _{0}\partial _{0}$ with $d\varsigma _{0}=\frac{d\xi }{\xi }$ and
$\partial _{0}$ as before. Notice moreover that one can express the field
action $\mathcal{S}_{2}\left[ \mathcal{X},\mathcal{Y},\mathcal{\Lambda }%
\right] $ in term of the $\mathcal{C}_{\pm }$ gauge field fluctuations.
Using the splitting $\mathcal{X}^{i}=x^{i}+\mu \mathcal{C}_{+}^{i}$ and $%
\mathcal{Y}_{i}=y_{i}-\mu \mathcal{C}_{-i}$, we obtain%
\begin{equation}
\mathcal{L}_{2}\left( \mathcal{C}_{\pm },\Lambda \right) =\frac{\mathrm{B}%
\mu }{2i}\left[ \left( \mathcal{C}_{-i}\partial _{0}\mathcal{C}_{+}^{i}-%
\mathcal{C}_{+}^{i}\partial _{0}\mathcal{C}_{-i}\right) +i\Lambda \left(
y_{i}\mathcal{C}_{+}^{i}-\mathcal{C}_{-i}x^{i}-\mathcal{C}_{-i}\mathcal{C}%
_{+}^{i}\right) \right] ,  \label{d}
\end{equation}%
where we have dropped out the total derivatives $\frac{d}{d\xi }\left( y_{i}%
\mathcal{C}_{+}^{i}+x^{i}\mathcal{C}_{-i}\right) $. Doing the same thing for
the splitting $\mathcal{X}^{i}=x^{i}E_{+-}+y^{i}A_{++}$ and $\mathcal{Y}%
_{i}=y_{i}E_{-+}-x_{i}A_{--}$ and substituting these relations back into eq(%
\ref{30}), we get,%
\begin{eqnarray}
\mathcal{L}_{2}\left[ E,A,\widetilde{\Lambda }\right] &=&\frac{\mu ^{2}%
\mathrm{B}}{2}\left( E_{-+}\partial _{0}E_{+-}-E_{+-}\partial
_{0}E_{-+}\right)  \notag \\
&&\frac{\mu ^{2}\mathrm{B}}{2}\left( A_{--}\partial
_{0}A_{++}-A_{++}\partial _{0}A_{--}\right)  \label{31} \\
&&+\frac{\mu \mathrm{B}}{2}\widetilde{\Lambda }\left(
E_{+-}E_{-+}-A_{++}A_{--}-1\right) ,  \notag
\end{eqnarray}%
invariant under the projective symmetry with $\widetilde{\Lambda }$ a
Lagrange gauge field parameter carrying the conifold constraint
hypersurface. By using the $D_{0}$ charge operator, the transformations (\ref%
{pr}) can be also stated as $D_{0}E_{\pm \mp }=0,$ $D_{0}A_{\pm \pm }=\pm
2A_{\pm \pm }$; they follow as well from the identities $D_{0}\mathcal{X}%
^{i}=\mathcal{X}^{i}$ and $D_{0}\mathcal{Y}^{i}=-\mathcal{Y}^{i}$. Note that
by substituting $E_{+-}=1+A_{+-}$ and $E_{-+}=1-A_{+-}$, one sees that the
term $\left( E_{-+}\partial _{0}E_{+-}-E_{+-}\partial _{0}E_{-+}\right) $
reduces to a total derivative $\partial _{0}\left( 2A_{+-}\right) $ and so
can be ignored in such a realization.

\subsection{Implementing density constraint equation}

\qquad First note that to get the density constraint eq in the fluid
approximation, one computes the total number $N$ of D strings by using two
paths; once by the coordinate frame $\left\{ x,y\right\} $ and second by
using the frame $\left\{ \mathcal{X},\mathcal{Y}\right\} $. Then equating
the two expressions since this number is invariant under coordinate
transformation. Supposing that fluid density is uniform $\rho \left( \xi
,x,y\right) =\rho _{0}$, a property implying,%
\begin{equation}
N=\int_{T^{\ast }S^{3}}\rho d^{3}v=\rho _{0}\int_{T^{\ast }S^{3}}d^{3}v,
\label{33}
\end{equation}%
and using the fact that this number is a constant, one gets a constraint eq
on the Jacobian $J\left( x,y\right) =\left\vert \frac{\partial ^{2}\left(
\mathcal{X},\mathcal{Y}\right) }{\partial ^{2}\left( x,y\right) }\right\vert
$ of the general transformation,%
\begin{equation}
x\qquad \rightarrow \qquad \mathcal{X}=\mathcal{X}\left( x,y\right) ,\qquad
y\qquad \rightarrow \qquad \mathcal{Y}=\mathcal{Y}\left( x,y\right) .
\label{34}
\end{equation}%
Eq(\ref{33}) requires that $J\left( x,y\right) =1$. Let us give some details
on this calculation. Since the density is uniform, we should have%
\begin{equation}
\rho _{0}\int_{T^{\ast }S^{3}}d^{3}\mathcal{V}=\rho _{0}\int_{T^{\ast
}S^{3}}d^{3}v.  \label{35}
\end{equation}%
Using the explicit expressions of the conifold holomorphic volume 3-form
which we write first as $d^{3}\mathcal{V}=\frac{d\xi }{\xi }\wedge d^{2}S$
and second $d^{3}v=\frac{d\xi }{\xi }\wedge d^{2}s$. Then expanding the K3
holomorphic 2-form $d^{2}S=\left( d\mathcal{X}^{i}\wedge d\mathcal{Y}%
_{i}\right) $, we get after some straightforward algebra,%
\begin{eqnarray}
\mu d^{2}S &=&\left\{ \mathcal{X}^{i},\mathcal{Y}_{i}\right\} _{+-}d^{2}s
\label{36} \\
&&+\left\{ \mathcal{X}^{i},\mathcal{Y}_{i}\right\} _{0-}dx^{l}\wedge
dx_{l}+\left\{ \mathcal{X}^{i},\mathcal{Y}_{i}\right\} _{0+}dy^{l}\wedge
dy_{l}.  \notag
\end{eqnarray}%
In this relation $d^{2}s=\left( dx^{i}\wedge dy_{i}\right) $ and $\left\{
f,g\right\} _{p,q}$ stand for the Poisson brackets defined as,%
\begin{eqnarray}
\left\{ f,g\right\} _{+-} &=&\left( D_{++}f\right) \left( D_{--}g\right)
-\left( D_{--}f\right) \left( D_{++}g\right) ,  \notag \\
\left\{ f,g\right\} _{0-} &=&\left( D_{0}f\right) \left( D_{--}g\right)
-\left( D_{--}f\right) \left( D_{0}g\right) ,  \label{37} \\
\left\{ f,g\right\} _{0+} &=&\left( D_{0}f\right) \left( D_{++}g\right)
-\left( D_{++}f\right) \left( D_{0}g\right) ,  \notag
\end{eqnarray}%
with $D_{\pm \pm ,0}$ generating the $SL\left( 2,C\right) $ isometry eqs(\ref%
{38}). Volume preserving diffeomorphisms require then the following
constraint eqs to be hold,%
\begin{equation}
\left\{ \mathcal{X}^{i},\mathcal{Y}_{i}\right\} _{+-}=\left( D_{++}\mathcal{X%
}^{i}\right) \left( D_{--}\mathcal{Y}_{i}\right) -\left( D_{--}\mathcal{X}%
^{i}\right) \left( D_{++}\mathcal{Y}_{i}\right) =\mu ,  \label{39}
\end{equation}%
and
\begin{eqnarray}
\left\{ \mathcal{X}^{i},\mathcal{Y}_{i}\right\} _{0-} &=&\left( D_{0}%
\mathcal{X}^{i}\right) \left( D_{--}\mathcal{Y}_{i}\right) -\left( D_{--}%
\mathcal{X}^{i}\right) \left( D_{0}\mathcal{Y}_{i}\right) =0,  \label{40} \\
\left\{ \mathcal{X}^{i},\mathcal{Y}_{i}\right\} _{0+} &=&\left( D_{0}%
\mathcal{X}^{i}\right) \left( D_{++}\mathcal{Y}_{i}\right) -\left( D_{++}%
\mathcal{X}^{i}\right) \left( D_{0}\mathcal{Y}_{i}\right) =0.  \notag
\end{eqnarray}%
A careful inspection shows that the last two conditions are not really
constraint eqs. The point is that because of the identities,%
\begin{equation}
D_{0}\mathcal{X}^{i}=\mathcal{X}^{i},\qquad D_{0}\mathcal{Y}_{i}=-\mathcal{Y}%
_{i},  \label{41}
\end{equation}%
required by K3 geometry, the two last constraint eqs can be brought to,%
\begin{equation}
\left\{ \mathcal{X}^{i},\mathcal{Y}_{i}\right\} _{0-}=D_{--}\left( \mathcal{X%
}^{i}\mathcal{Y}_{i}\right) ,\qquad \left\{ \mathcal{X}^{i},\mathcal{Y}%
_{i}\right\} _{0+}=D_{++}\left( \mathcal{X}^{i}\mathcal{Y}_{i}\right) .
\label{42}
\end{equation}%
But these relations vanishes identically because of the identity $\mathcal{X}%
^{i}\mathcal{Y}_{i}=\mu =$ {\small constant}. Therefore the volume
transformation (\ref{36}) becomes $\mu d^{2}S=\left\{ \mathcal{X}^{i},%
\mathcal{Y}_{i}\right\} _{+-}d^{2}s$ and so we are left with one constraint
relation; namely $\left\{ \mathcal{X}^{i},\mathcal{Y}_{i}\right\} _{+-}=\mu $
which can be implemented in the field action (\ref{30}) by help of a
Lagrange gauge field $\mathcal{C}_{0}$. To that purpose note that by setting
$\mathcal{J}_{\pm \pm }=\pm \left( \mathcal{C}_{0}\mathcal{Y}_{i}D_{\pm \pm }%
\mathcal{X}^{i}\right) $, one can check that we have,%
\begin{equation}
\int d^{3}v\mathcal{C}_{0}\left[ \left\{ \mathcal{X}^{i},\mathcal{Y}%
_{i}\right\} _{+-}-\mu \right] =\int d^{3}v\left( \mathcal{Y}_{i}\left\{
\mathcal{C}_{0},\mathcal{X}^{i}\right\} _{+-}\right)  \label{43}
\end{equation}%
where we have dropped out the boundary term $\int d^{3}v\left[ D_{--}%
\mathcal{J}_{++}+D_{++}\mathcal{J}_{--}\right] $. Implementing this identity
in the field action as usual, we get the following holomorphic functional%
\begin{equation}
\mathcal{S}_{DSF}\left[ \mathcal{X},\mathcal{Y},\mathcal{C}_{0}\right]
=\int_{T^{\ast }S^{3}}d^{3}v\text{ }\mathcal{L}_{DSF}\left( \mathcal{X},%
\mathcal{Y},\mathcal{C}_{0}\right) ,
\end{equation}%
with,%
\begin{eqnarray}
\mathcal{L}_{DSF}\left[ \mathcal{X},\mathcal{Y},\mathcal{C}_{0}\right] &=&%
\frac{i\mathrm{B}}{2\mu }\left( \mathcal{Y}_{i}\partial _{0}\mathcal{X}^{i}-%
\mathcal{X}^{i}\partial _{0}\mathcal{Y}_{i}\right) +\frac{\mathrm{B}}{2\mu }%
\Lambda \left( \mathcal{Y}_{i}\mathcal{X}^{i}-\mu \right)  \notag \\
&&-\frac{\mathrm{B}}{\mu }\left( \mathcal{Y}_{i}\left\{ \mathcal{C}_{0},%
\mathcal{X}^{i}\right\} _{+-}-\mathcal{X}^{i}\left\{ \mathcal{C}_{0},%
\mathcal{Y}_{i}\right\} _{+-}\right) .  \label{44}
\end{eqnarray}%
Using the previous splitting of the D string fields $\mathcal{X}^{i}$ and $%
\mathcal{Y}_{i}$, we can express this field action in terms of the gauge
fields either as $\mathcal{S}_{DSF}=\mathcal{S}_{DSF}\left[ \mathcal{C}_{\pm
i},\mathcal{C}_{0},\Lambda \right] $ or equivalently as $\mathcal{S}_{DSF}=%
\mathcal{S}_{DSF}\left[ E,A,\mathcal{C}_{0},\Lambda \right] $. Let us do
this calculation for the splitting $\mathcal{X}^{i}=x^{i}+\mu \mathcal{C}%
_{+}^{i}$ and $\mathcal{Y}_{i}=y_{i}-\mu \mathcal{C}_{-i}$. In this case the
density constraint eq $\left\{ \mathcal{X}^{i},\mathcal{Y}_{i}\right\}
_{+-}=\mu $ reads in terms of the $\mathcal{C}_{\pm i}$ gauge fields as
follows,
\begin{equation}
\left\{ x^{i},\mathcal{C}_{-i}\right\} _{-+}-\left\{ \mathcal{C}%
_{+}^{i},y_{i}\right\} _{-+}+i\mu \left\{ \mathcal{C}_{+}^{i},\mathcal{C}%
_{-i}\right\} _{-+}=0.  \label{420}
\end{equation}%
This relation can be put into a more interesting way by setting $\left\{
x^{i},F\right\} _{-+}=\partial _{+}^{i}F$, $\left\{ F,y_{i}\right\}
_{-+}=\partial _{-i}F$\ with the remarkable properties $\partial
_{+}^{i}\partial _{-i}=-y_{i}D_{++}\left( x^{i}D_{--}\right) =-\mu
D_{++}D_{--}$ and $\partial _{-i}\partial _{+}^{i}=-x^{i}D_{--}\left(
y_{i}D_{++}\right) =-\mu D_{--}D_{++}$. Putting these relations back into (%
\ref{420}), we obtain%
\begin{equation}
\partial _{+}^{i}\mathcal{C}_{-i}-\partial _{-i}\mathcal{C}_{+}^{i}-i\left(
\partial _{+}^{k}\mathcal{C}_{+}^{i}\partial _{-k}\mathcal{C}_{-i}-\partial
_{-k}\mathcal{C}_{+}^{i}\partial _{+}^{k}\mathcal{C}_{-i}\right) =0,
\label{421}
\end{equation}%
or equivalently by introducing Poisson bracket $\left\{ F,G\right\}
_{PB}\equiv \left( \partial _{+k}F\right) \left( \partial _{-}^{k}G\right)
-\left( \partial _{-}^{k}F\right) \left( \partial _{+k}G\right) $,%
\begin{equation}
\partial _{+}^{i}\mathcal{C}_{-i}-\partial _{-i}\mathcal{C}_{+}^{i}-i\left\{
\mathcal{C}_{+}^{i},\mathcal{C}_{-i}\right\} _{PB}=0.  \label{422}
\end{equation}%
Note also that $\left\{ F,G\right\} _{PB}$ is just $\mu \left\{ F,G\right\}
_{-+}$. As we see, this is a typical equation of motion of non commutative
gauge theory; it can be then thought of as the minimization of an invariant
gauge field $\mathcal{S}_{DSF}\left[ \mathcal{C}_{\pm },\mathcal{C}_{0}%
\right] $ with gauge fields $\mathcal{C}_{\pm }^{i}$ and $\mathcal{C}_{0}$.
In this view, we have,
\begin{equation}
\frac{\delta \mathcal{S}_{DSF}\left[ \mathcal{C}_{\pm },\mathcal{C}_{0}%
\right] }{\delta \mathcal{C}_{0}}=\partial _{+}^{i}\mathcal{C}_{-i}-\partial
_{-i}\mathcal{C}_{+}^{i}-i\left\{ \mathcal{C}_{+}^{i},\mathcal{C}%
_{-i}\right\} _{PB}=0,  \label{423}
\end{equation}%
from which we can determine $\mathcal{S}_{DSF}\left[ \mathcal{C}_{\pm },%
\mathcal{C}_{0}\right] $ taking into account eq(\ref{d}). Setting
\begin{equation*}
\mathcal{S}_{DSF}\left[ \mathcal{C}_{\pm }^{i},\mathcal{C}_{0},\Lambda %
\right] =\frac{i\mathrm{B}}{2\mu }\int_{T^{\ast }S^{3}}d^{3}v\mathcal{L}%
_{DSF}\left[ \mathcal{C}_{\pm },\mathcal{C}_{0},\Lambda \right] ,
\end{equation*}%
we have
\begin{eqnarray}
\mathcal{L}_{DFS}\left[ \mathcal{C}_{\pm },\mathcal{C}_{0},\Lambda \right]
&=&\frac{i\mathrm{B}\mu }{2}\left( \mathcal{C}_{+}^{i}\partial _{0}\mathcal{C%
}_{-i}-\mathcal{C}_{-i}\partial _{0}\mathcal{C}_{+}^{i}\right) +  \notag \\
&&-\frac{\mathrm{B}\mu }{2}\left[ 2\left( \mathcal{C}_{0}\partial _{+}^{i}%
\mathcal{C}_{-i}-\mathcal{C}_{0}\partial _{-i}\mathcal{C}_{+}^{i}\right) -2%
\mathcal{C}_{0}\left\{ \mathcal{C}_{+}^{i},\mathcal{C}_{-i}\right\} _{PB}%
\right] \\
&&+\frac{\mathrm{B}\mu }{2}\Lambda \left( y_{i}\mathcal{C}_{+}^{i}-\mathcal{C%
}_{-i}x^{i}-\mathcal{C}_{-i}\mathcal{C}_{+}^{i}\right) .  \notag
\end{eqnarray}%
This holomorphic lagrangian density may be put into a more convenient way by
performing an integration by part and dropping out the total derivatives.
Replacing%
\begin{equation}
\mathcal{C}_{0}\left\{ F,G\right\} _{-+}=-F\left\{ \mathcal{C}_{0},G\right\}
_{-+}+F\mathcal{C}_{0}D_{0}G+\text{ {\small total derivative}}
\end{equation}%
for holomorphic functions $F$ and $G$ on conifold we have,%
\begin{eqnarray}
\mathcal{L}_{DSF}\left[ \mathcal{C}_{\pm },\mathcal{C}_{0}\right] &=&\frac{i%
\mathrm{B}\mu }{2}\left[ \mathcal{C}_{+}^{i}\partial _{0}\mathcal{C}_{-i}-%
\mathcal{C}_{+}^{i}\partial _{-i}\mathcal{C}_{0}-\frac{2i}{3}\mathcal{C}%
_{+}^{i}\left\{ \mathcal{C}_{0},\mathcal{C}_{-i}\right\} _{PB}\right]  \notag
\\
&&+\frac{i\mathrm{B}\mu }{2}\left[ -\mathcal{C}_{0}\partial _{+}^{i}\mathcal{%
C}_{-i}+\mathcal{C}_{0}\partial _{-i}\mathcal{C}_{+}^{i}+\frac{2i}{3}%
\mathcal{C}_{0}\left\{ \mathcal{C}_{+}^{i},\mathcal{C}_{-i}\right\} _{PB}%
\right]  \label{hcs} \\
&&+\frac{i\mathrm{B}\mu }{2}\left[ \mathcal{C}_{-i}\partial _{+}^{i}\mathcal{%
C}_{0}-\mathcal{C}_{-i}\partial _{0}\mathcal{C}_{+}^{i}-\frac{2i}{3}\mathcal{%
C}_{-i}\left\{ \mathcal{C}_{+}^{i},\mathcal{C}_{0}\right\} _{PB}\right] .
\notag
\end{eqnarray}%
where we have set $\left( y_{i}\mathcal{C}_{+}^{i}-x^{i}\mathcal{C}_{-i}-%
\mathcal{C}_{+}^{i}\mathcal{C}_{-i}\right) =0$ describing gauge fluctuations
restricted to conifold. By substituting $\mu \mathcal{C}%
_{+}^{i}=x^{i}A_{+-}+y^{i}A_{++}$ and $\mu \mathcal{C}%
_{-i}=-y_{i}A_{-+}+x_{i}A_{--}$ in the above gauge field action, one gets
the expression of $\mathcal{L}_{DSF}\left[ A_{+-},A_{++},A_{-+},A_{--}\right]
$\ in terms of the gauge fields $A_{+-},$ $A_{++},$ $A_{-+}$ and $A_{--}$ .

\qquad In the end notice that on the real slice of conifold with parameter $%
\func{Re}\mu $, background field $\func{Re}\mathrm{B}$ and field variables
as,
\begin{equation}
\mathcal{Y}_{i}=\overline{\left( \mathcal{X}^{i}\right) },\qquad
\Leftrightarrow \qquad \mathcal{C}_{-i}=\overline{\left( \mathcal{C}%
_{+}^{i}\right) },
\end{equation}%
the previous field action reduces to non commutative Chern Simons gauge
theory in real three dimensions. In this case $\left( \func{Re}\mathrm{B}%
\right) \times \left( \func{Re}\mu \right) $ should be equal to Kac-Moody
level $k$.

\section{Holomorphy and quantum corrections}

\qquad Though natural from classical view, the correspondence between FQH
systems and fluids of D-strings in conifold described above is however no
longer obvious at quantum level. In the D-string fluid proposal, the
classical\ free degrees of freedom of the holomorphic sector,%
\begin{equation}
\mathcal{S}_{N}\left[ X,Y\right] =\frac{i\mathrm{B}}{2}\int_{T^{\ast
}S^{1}}d\zeta \sum_{a=1}^{N}\left( Y_{ia}\frac{\partial X_{a}^{i}}{\partial
\zeta }-X_{a}^{i}\frac{\partial Y_{ia}}{\partial \zeta }\right) ,
\end{equation}
and the corresponding antiholomorphic one,
\begin{equation}
\mathcal{S}_{N}^{\ast }\left[ X^{\ast },Y^{\ast }\right] =\frac{-i\mathrm{B}%
}{2}\int_{T^{\ast }S^{1}}d\overline{\zeta }\sum_{a=1}^{N}\left( Y_{ia}^{\ast
}\frac{\partial X_{a}^{\ast i}}{\partial \overline{\zeta }}-X_{a}^{\ast i}%
\frac{\partial Y_{ia}^{\ast }}{\partial \overline{\zeta }}\right) ,
\end{equation}%
may couple quantum mechanically unless this is forbidden by underlying
symmetries. Typical examples of these powerful symmetries, one encounters in
such kind of situations, are generally given by conformal invariance,
supersymmetry and their extensions. In this section, we make general
comments on quantum effects in the D string system and give a discussion on
how supersymmetry can help to overcome difficulties. Implication of
supersymmetry in the game can be motivated from several views starting from
complex Kahler geometry of $T^{\ast }\mathbb{S}^{3}$ and ending with
topological aspects of $2d$ fields on conifold. To fix the ideas on the way
we will do things, we recall the standard parallel between field holomorphy
in conifold geometry and chirality in $2d$ $\mathcal{N}=2$\ supersymmetric
non linear sigma model captured by the usual supersymmetric derivatives $%
\overline{D}_{\pm 1/2}$. Using this parallel, we shall show that the
holomorphic lagrangian density $L\left( X,Y\right) =\mathrm{B}%
\sum_{a=1}^{N}Y_{ia}\left( \partial X_{a}^{i}/\partial \zeta \right) $ of
the D-string fluid can be thought of as following from the chiral superspace
lagrangian of the $\mathcal{N}=2$ supersymmetric sigma model in large B
field,
\begin{equation}
L_{chiral}\left[ \Phi \right] =\int_{SM_{-}}d^{2}\mathrm{\theta }\text{ }%
\mathcal{W}\left( \Phi \right) ,
\end{equation}%
where $\Phi $ refers to generic chiral superfields and $SM_{-}$\ to chiral
superspace. In this relation, $\mathcal{W}\left( \Phi \right) \sim \left(
\mathrm{B}\sum_{a=1}^{N}\Phi _{a1}\Phi _{a2}\right) $ is chiral the
superpotential. Substituting the chiral superfields $\Phi _{ia}$ by their $%
\theta $-expansions; i.e
\begin{equation}
\Phi _{ia}\sim Y_{ia}+...+\theta {\small _{\text{+1/2}}}\theta {\small _{%
\text{-1/2}}}F_{ia},\qquad i=1,2,
\end{equation}%
where we have dropped out fermions and where $F_{ia}$ are auxiliary fields
to be specified in a moment; then integrating with respect to the Grassman
variables ${\small \theta }_{{\small \pm }\text{{\small 1/2}}}$, gives the
following field component product $\mathrm{B}\left(
\sum_{a=1}^{N}Y_{ia}F_{ia}\right) $. By taking the auxiliary fields $F_{ia}$%
\ as,%
\begin{equation}
F_{ia}=\left( \sqrt{\mu }\epsilon _{ij}X_{a}^{j}+\epsilon _{ij}\frac{%
\partial X_{a}^{j}}{\partial \zeta }\right) ,
\end{equation}%
where $\epsilon _{ij}$ is the usual spinor metric and $\mu $ the conifold
complex parameters, one discovers, up to a constant, the above holomorphic
lagrangian density.

\qquad Before going ahead, it should be also noted that the comments we
shall give below are certainly not final answers; but just a tentative to
approach aspects of quantum behaviour of D string fluid in conifold. The
discussion presented below relies on path integral method for quantization.
But may be the more natural way to do would be extending matrix model
approach of Susskind-Polychronakos (SP) for FQH droplets. Recall that SP
method uses canonical quantization. We will give a brief comment on this
method in the end of this section. More involved details may be found in
\cite{15}.

\qquad This discussion is organized as follows: In the first subsection, we
explore the consequences of quantum effects on conifold geometry and derive
the constraint eq on quantum consistency of holomorphy property. Using path
integrals quantization method, we show that holomorphy persists as far
quantum fluctuations are restricted to complex deformations of conifold.
Implementation of Kahler deformations destroys this behaviour since
holomorphic and antiholomorphic modes get coupled. In sub-section 2, we
study the embedding the D string model in a supersymmetric theory and too
particularly in its chiral sectors. The latter seems to be the appropriate
theory that governs the quantum fluctuations of the D-string fluid in
conifold. As a first step in checking this statement, we start by describing
the field theoretic derivation of holomorphy hypothesis considered in
section 2. Then we give a correspondence with $2d$ $\mathcal{N}=2$
supersymmetric non linear sigma model with conifold as a target space; in
presence of a background magnetic field $B$. We end this section by
discussing the statistics of the D-string system which requires a filling
fraction $\nu =\frac{1}{k}$ with even integer Kac-Moody levels $k$.

\subsection{Quantum effects and conifold deformations}

\qquad A way\footnote{%
An other tentative to approach the fluid of D-strings in conifold, by using
a generalization of matrix model method based on canonical quantization, has
been developed in \cite{15}. There and as a first step in dealing with the
problem, one focuses on the study of quantum droplets for the conifold
sub-varieties $\mathbb{S}^{3}$ and $\mathbb{S}^{2}$.} to study\ the quantum
effects on the holomorphy feature of the D-string fluid model is to proceed
as follows. First think about the D string fluid model as a classical field
theory based on the conifold geometry $xy-zw=\mu $. This means that the
complex threefold, with its complex modulus $\mu $, can be thought of as a
classical geometry. Quantum mechanically, the above fields are subject to
fluctuations and so the complex parameter $\mu $ gets corrections induced by
quantum effects. To have an idea on the nature of these quantum corrections,
we consider fluctuations of the D-strings around the classical field
configurations $x,y,z,w$. These field fluctuations can be written as%
\begin{equation}
\phi \qquad \rightarrow \qquad \phi +\delta F_{\phi },\qquad \phi =x,y,z,w,
\label{d1}
\end{equation}%
with the generic fields $\phi =\phi \left( \xi \right) $ is as in eq(\ref{5}%
) and $\delta F_{\phi }$ describing the perturbations around the classical
field $\phi $. Notice that these fluctuations are involved in the
computation of the partition function $\mathcal{Z}\left[ j\right] $ of the
model,%
\begin{equation}
\mathcal{Z}\left[ j\right] =\int \left( \dprod D\phi \right) \exp \frac{i}{%
\hbar }\left( S\left[ \phi \right] +\int j\phi \right) ,
\end{equation}%
where $\dprod D\phi $ stays for the usual field path integral measure. As it
is known, this quantity generates the Green functions of the quantum model
with $j$ being the usual external source. Notice also that the $\delta
F_{\phi }$ deformations should a priori depend on both the string fields $%
\phi $ and their complex conjugates $\overline{\phi }$ as shown below,
\begin{equation}
\delta F_{\phi }=F\left( \phi ,\overline{\phi }\right) .  \label{d2}
\end{equation}%
By implementing the fluctuations (\ref{d1}) into the D-string fluid model,
one discovers that the classical geometry $xy-zw=\mu $ we started with gets
now deformed as follows,
\begin{equation}
xy-zw=\mu +\mathcal{F},
\end{equation}%
where the functional $\mathcal{F}$ capturing the field fluctuations is given
by,
\begin{equation}
\mathcal{F}=x\delta F_{y}+y\delta F_{x}-z\delta F_{w}-w\delta F_{z}.
\end{equation}%
Like for eq(\ref{d2}), one sees that $\mathcal{F}$ depends in general on
both the fields $x$, $y$, $z$, $w$ and their complex conjugate $\overline{x}$%
, $\overline{y}$, $\overline{z}$, $\overline{w}$,%
\begin{equation}
\mathcal{F}=\mathcal{F}\left( \phi ,\overline{\phi }\right) ,\qquad \phi
=x,y,z,w.
\end{equation}%
Thus, quantum mechanical effects encoded in the functional $\mathcal{F}%
\left( \phi ,\overline{\phi }\right) $ break holomorphy of the classical
conifold geometry unless field deformations $\delta F_{\phi }$ are
restricted to holomorphic perturbations around the classical field
configuration. In this special case, we have,%
\begin{equation}
\frac{\partial \mathcal{F}}{\partial \overline{\phi }}=0,  \label{d6}
\end{equation}%
and so classical holomorphy is preserved quantum mechanically. This is the
condition for quantum decoupling of holomorphic and antiholomorphic degrees
of freedom. This property has a geometric interpretation in term of conifold
structure deformations; it means that only complex deformations of
holomorphic volume that are allowed for having a consistent quantum
mechanics. It is also interesting to note that eq(\ref{d6}) is a strong
condition; its solution requires however a strong symmetry which apparently
D-string fluid model does not exhibit manifestly; at least not as things
have been formulated so far. Note moreover that as far as quantum holomorphy
is concerned, to our knowledge only supersymmetry that has the magic power
to deal with target space holomorphy. There, quantum corrections are
controlled by the so called non renormalization theorem.

\qquad The next question is how the string fluid model could be related to $%
2d$ $\mathcal{N}=2$ supersymmetric non linear sigma model with conifold as
target space. Thinking about the D-string model as the bosonic part of a
supersymmetric theory does not answer exactly the question since there are
Kahler deformations induced by quantum effects that destroy the classical
holomorphy property. To overcome such difficulty one should then associate
the action of the D-string model with chiral superpotentials,%
\begin{equation}
\mathcal{W}=\mathcal{W}\left( \Phi \right) ,
\end{equation}%
of $\mathcal{N}=2$ supersymmetric non linear sigma model. In what follows,
we develop a way to do it. Though not exact and needs more investigations,
this approach offers however an important step towards the goal.

\subsection{Supersymmetric embedding}

\qquad To begin recall that there is a closed connection between Kahler
geometry and $\mathcal{N}=2$ supersymmetry in two dimensions. The fact that
the fluid of D-strings is described by a topological holomorphic gauge
theory, let understand that this model can be embedded in a $\mathcal{N}=2$
supersymmetric theory; from which one can get informations about quantum
corrections. In this view holomorphy property is interpreted as the target
space manifestation of chirality feature of $2d$ $\mathcal{N}=2$
supersymmetric sigma models with conifold as target space. A close idea is
used in building topological string theory by using twist of $2d$ $N=2$
superconformal algebra \cite{17} and a correspondence with type II
superstrings on Calabi-Yau threefolds \cite{18}. In our concern, we have the
following correspondence,%
\begin{eqnarray}
\int d\zeta ...\qquad &\rightarrow &\qquad \int d^{2}\mathrm{\theta }...,
\notag \\
\int d\overline{\zeta }...\qquad &\rightarrow &\qquad \int d^{2}\overline{%
\mathrm{\theta }}..., \\
\int d^{2}\zeta ...\qquad &\rightarrow &\qquad \int d^{2}\mathrm{\theta }%
d^{2}\overline{\mathrm{\theta }}...,  \notag
\end{eqnarray}%
with the $\mathrm{\theta }_{\pm 1/2}$'s and $\overline{\mathrm{\theta }}%
_{\pm 1/2}$'s the usual Grassman variables. Similar things may be also
written down for $\partial /\partial \zeta $ and supersymmetric derivatives.
Before that, let us start by deriving rigorously the holomorphy hypothesis
of section 2 by using a field theoretical method; then come back to the
correspondence between target space holomorphy and 2d $\mathcal{N}=2$
supersymmetric chirality.

\subsubsection{Holomorphy property and boundary QFT$_{2}$}

\qquad Holomorphy is one of the basic ingredients we have used in deriving
the D-string model developed in this paper. It has been imposed in order to
complete the conifold realization $T^{\ast }\mathbb{S}^{3}$ as a fibration
of $T^{\ast }\mathbb{S}^{1}$ over the base $T^{\ast }\mathbb{S}^{2}$. In
this study, we first give the field theoretic derivation of this holomorphy
hypothesis; it appears as the solution of a constraint eq required by
boundary field theory in two dimensions. Then we derive the field action (%
\ref{27}); its connection with supersymmetric models is considered in the
next sub-subsection.

\qquad To proceed and seen that the model we are studying involves complex
fields, it is then natural to start from the following bosonic QFT$_{2}$
field action,%
\begin{equation}
S\left[ \phi ,\overline{\phi }\right] =\int_{M}d^{2}\zeta \left( G_{\alpha
\overline{\beta }}\partial _{+}\phi ^{\alpha }\partial _{-}\overline{\phi
^{\beta }}\right) ,  \label{d7}
\end{equation}%
where $M$ is a real surface parameterized by the local complex coordinates $%
\left( \zeta ,\overline{\zeta }\right) $. The fields $\phi ^{\alpha }=\phi
^{\alpha }\left( \zeta ,\overline{\zeta }\right) $ form a set of complex $2d$
scalar fields parameterizing some target Kahler manifold with metric $%
G_{\alpha \overline{\beta }}=G_{\alpha \overline{\beta }}\left( \phi ,%
\overline{\phi }\right) $. To make contact with conifold geometry and the
fluid of N strings, we think about these field variables as,
\begin{equation}
\phi ^{\alpha }\left( \zeta ,\overline{\zeta }\right) =X_{a}^{i}\left( \zeta
,\overline{\zeta }\right) ,\qquad i=1,2,\qquad a=1,...,N,
\end{equation}%
with $X_{a}^{i}$\ an SU$\left( 2\right) $ doublet like in eq(\ref{11}) and
to fix the ideas the field doublet $Y_{ia}$ are set to $\overline{X_{a}^{i}}$%
. Once the idea is exhibited, the field $\overline{X_{a}^{i}}$ will be
promoted to $Y_{ia}$. In this case, the Kahler metric $G_{\alpha \overline{%
\beta }}$ may be split as
\begin{equation}
G_{\alpha \overline{\beta }}\left( \phi ,\overline{\phi }\right) =\delta
_{ab}\left[ g_{\left( ij\right) }+B\epsilon _{ij}\right] ,
\end{equation}%
where the $SU\left( 2\right) $ triplet $g_{\left( ij\right) }$ is a function
on the target space field coordinates; i.e $g_{\left( ij\right) }=g_{\left(
ij\right) }\left( \phi ,\overline{\phi }\right) $, and where $\epsilon _{ij}$
is the usual antisymmetric $SU\left( 2\right) $ invariant tensor. In the
special case where $B$ is field independent and strong enough so that we can
neglect the term $g_{\left( ij\right) }$, the metric $G_{\alpha \overline{%
\beta }}$ reduces essentially to $B\delta _{ab}\epsilon _{ij}$; and so one
is left with the following approximated field action,%
\begin{equation}
S\left[ X,\overline{X}\right] \simeq \int_{M}d^{2}\zeta \left( B\epsilon
_{ij}\sum_{a=1}^{N}\partial _{+}\overline{X}_{a}^{j}\partial
_{-}X_{a}^{i}\right) ,  \label{d100}
\end{equation}%
where we have set $\zeta =\zeta _{-}$, $\overline{\zeta }=\zeta _{+}$ and $%
\partial _{+}=\partial _{\zeta }$, $\partial _{-}=\partial _{\overline{\zeta
}}$. Moreover since $B$ is a constant, one can split this action as follows,
\begin{eqnarray}
S\left[ X,\overline{X}\right] &\simeq &\frac{B}{2}\int_{M}d\zeta _{-}\left[
d\zeta _{+}\partial _{-}\left( \sum_{a=1}^{N}\left( \partial _{+}\overline{X}%
_{ia}\right) X_{a}^{i}\right) \right]  \notag \\
&&+\frac{B}{2}\int_{M}d\zeta _{+}\left[ d\zeta _{-}\partial _{+}\left(
\sum_{a=1}^{N}\overline{X}_{ia}\left( \partial _{-}X_{a}^{i}\right) \right) %
\right]  \label{d11} \\
&&-\frac{B}{2}\int_{M}d^{2}\zeta \sum_{a=1}^{N}\left[ \left( \partial
_{-}\partial _{+}\overline{X}_{ia}\right) X_{a}^{i}+\overline{X}_{ia}\left(
\partial _{+}\partial _{-}X_{a}^{i}\right) \right] ,  \notag
\end{eqnarray}%
where the summation over $SU\left( 2\right) $ indices is understood. By
integrating the two first terms of above relation, one sees that the field
action $S\left[ X,\overline{X}\right] $ decomposes as,
\begin{equation}
S\simeq S^{{\tiny boundary}}+S^{{\tiny bulk}}  \label{d110}
\end{equation}%
with two factors for $S^{{\tiny boundary}}=S^{{\tiny bound}}$ as given
below,
\begin{eqnarray}
S^{{\tiny bound}} &=&\frac{B}{2}\int_{\partial M_{-}}d\zeta \left(
\sum_{a=1}^{N}\left( \partial _{+}\overline{X}_{ia}\right) X_{a}^{i}\right)
\notag \\
&&+\frac{B}{2}\int_{\partial M_{+}}d\overline{\zeta }\left( \sum_{a=1}^{N}%
\overline{X}_{ia}\left( \partial _{-}X_{a}^{i}\right) \right) ,  \label{d12}
\end{eqnarray}%
where $\partial M_{\pm }$ stand for the oriented boundaries of the Riemann
surface $M$ and
\begin{equation}
S^{{\tiny bulk}}=-\int_{M}d^{2}\zeta \sum_{a=1}^{N}\left[ \frac{B}{2}\left(
\partial _{-}\partial _{+}\overline{X}_{ia}\right) X_{a}^{i}+\frac{B}{2}%
\overline{X}_{ia}\left( \partial _{+}\partial _{-}X_{a}^{i}\right) \right] .
\label{d13}
\end{equation}%
Equating eq(\ref{d100}) and eq(\ref{d12}), one gets the holomorphy condition
of the field variables,%
\begin{equation}
\left[ \frac{\partial }{\partial \zeta }\frac{\partial }{\partial \overline{%
\zeta }}X_{a}^{i}\left( \zeta ,\overline{\zeta }\right) \right] _{\partial
M}=0,\qquad \left[ \frac{\partial }{\partial \zeta }\frac{\partial }{%
\partial \overline{\zeta }}\overline{X}_{ia}\left( \zeta ,\overline{\zeta }%
\right) \right] _{\partial M}=0.
\end{equation}%
These constraint relations are solved by field holomorphy as shown below;%
\begin{eqnarray}
X_{a}^{i}\left( \zeta ,\overline{\zeta }\right) &=&X_{a}^{i}\left( \zeta
\right) +X_{a}^{i}\left( \overline{\zeta }\right) ,  \notag \\
\overline{X}_{ia}\left( \zeta ,\overline{\zeta }\right) &=&X_{ia}^{\ast
}\left( \zeta \right) +X_{ia}^{\ast }\left( \overline{\zeta }\right) .
\end{eqnarray}%
They tell us that on the boundary $\partial M$ of the Riemann surface, we
have two heterotic free field theories; a holomorphic sector with field
variables%
\begin{equation}
X_{a}^{i}\left( \zeta \right) ,\qquad X_{ia}^{\ast }\left( \zeta \right) ,
\label{d16}
\end{equation}%
which, for convenience and avoiding confusion we set $X_{ia}^{\ast }\left(
\zeta \right) =Y_{ia}\left( \zeta \right) $, and an antiholomorphic one with,%
\begin{equation}
X_{ia}^{\ast }\left( \overline{\zeta }\right) =\overline{\left(
X_{a}^{i}\left( \zeta \right) \right) },\qquad X_{a}^{i}\left( \overline{%
\zeta }\right) =\overline{\left( Y_{ia}\left( \zeta \right) \right) },
\label{d17}
\end{equation}%
in agreement with the hypothesis on holomorphicity of the string coordinates.

\subsubsection{Supersymmetric interpretation}

\qquad The decomposition of the field action $S\left[ {\small QFT}_{2}\right]
$ eqs(\ref{d7}-\ref{d100}) taken in the limit large $B$ field is very
suggestive. First, because it explains the origin of holomorphy hypothesis
we have used to build the model of the fluid of D strings. As such, one
should distinguish between fields in bulk and fields in boundary given by
eqs(\ref{d16}-\ref{d17}). Second it permits a one to one correspondence with
$2d$ $\mathcal{N}=2$ supersymmetric non linear sigma models. More precisely
the three terms of the field action of the bosonic QFT$_{2}$ in large B
limit,%
\begin{eqnarray}
S\left[ {\small QFT}_{2}\right] &=&-\int_{M}d^{2}\zeta \sum_{a=1}^{N}\left[
\frac{B}{2}X_{a}^{i}\left( \partial _{-}\partial _{+}\overline{X}%
_{ia}\right) +\frac{B}{2}\overline{X}_{ia}\left( \partial _{+}\partial
_{-}X_{a}^{i}\right) \right]  \notag \\
&&-\int_{\partial M_{-}}d\zeta \left( \frac{B}{2}\sum_{a=1}^{N}Y_{ia}\left(
\partial _{+}X_{a}^{i}\right) \right)  \label{d15} \\
&&-\int_{\partial M_{+}}d\overline{\zeta }\left( \frac{B}{2}\sum_{a=1}^{N}%
\overline{\left( Y_{ia}\right) }\left( \partial _{-}\overline{\left(
X_{a}^{i}\right) }\right) \right) ,  \notag
\end{eqnarray}%
are in one to one with the usual three blocks of $2d$ $\mathcal{N}=2$
supersymmetric non linear sigma models,%
\begin{eqnarray}
S_{\mathcal{N}=2}\left[ \Phi ,\Phi ^{+}\right] &=&\int_{SM}d^{2}\upsilon
d^{2}\mathrm{\theta }d^{2}\overline{\mathrm{\theta }}\text{ }\mathcal{K}%
\left( \Phi _{i},\Phi _{i}^{+}\right)  \notag \\
&&+\int_{SM_{-}}d^{2}\upsilon d^{2}\mathrm{\theta }\text{ }\mathcal{W}\left(
\Phi _{i}\right) \\
&&+\int_{SM_{+}}d^{2}\upsilon d^{2}\overline{\mathrm{\theta }}\text{ }%
\overline{\mathcal{W}}\left( \Phi _{i}^{+}\right) .  \notag
\end{eqnarray}%
In this relation, the symbol $SM$ stands for the usual two dimensional
superspace with super-coordinates $\left( \upsilon _{\pm },\mathrm{\theta }_{%
{\small \pm }\text{{\small 1/2}}},\overline{\mathrm{\theta }}_{{\small \pm }%
\text{{\small 1/2}}}\right) $ and $SM_{\pm }$ stand for the two associated
chiral superspaces. The $\Phi _{i}$'s (resp. $\Phi _{i}^{+}$) are chiral
(resp. antichiral) superfields living on $SM_{-}$ (resp. $SM_{+}$), $%
\mathcal{K}\left( \Phi ,\Phi ^{+}\right) $ is the Kahler superpotential and $%
\mathcal{W}\left( \Phi \right) $ the usual complex chiral superpotential.
Like for the holomorphic functions $f=f\left( \zeta \right) $ living on $%
\partial M_{-}$ and satisfying the holomorphy property,
\begin{equation}
\frac{\partial f}{\partial \overline{\zeta }}=0,
\end{equation}%
we have for chiral superfields $\Phi \left( \widetilde{\upsilon }_{\pm },%
\mathrm{\theta }_{{\small \pm }\text{{\small 1/2}}}\right) $ living on $%
SM_{-}$, the following chirality property,%
\begin{equation}
\overline{D}_{\pm 1/2}\Phi =0.
\end{equation}%
By comparison of the two actions, one sees that the bulk term $S^{{\tiny bulk%
}}$ of the QFT$_{2}$ eq(\ref{d15}) is associated with Kahler term of the
supersymmetric sigma model,%
\begin{equation}
S^{{\tiny bulk}}\left[ \text{{\small QFT}}_{2}\right] \qquad
\longleftrightarrow \qquad \int_{SM}d^{2}\upsilon d^{2}\mathrm{\theta }d^{2}%
\overline{\mathrm{\theta }}\text{ }\mathcal{K}\left( \Phi _{i},\Phi
_{i}^{+}\right) ,
\end{equation}%
while the two boundary terms $S_{\pm }^{{\tiny bound}}$ are associated with
the chiral superfield actions. More precisely, we have
\begin{equation}
\int_{\partial M_{-}}d\zeta \left( \sum_{a=1}^{N}\frac{B}{2}%
Y_{ia}F_{a}^{i}\right) \qquad \longleftrightarrow \qquad
\int_{SM_{-}}d^{2}\upsilon d^{2}\mathrm{\theta }\text{ }\mathcal{W}\left(
\Phi _{i}\right) ,
\end{equation}%
where we have set $F_{a}^{i}=\left( \partial _{\zeta }X_{a}^{i}\right) $ and
by putting after setting $\overline{F}_{ia}=\left( \partial _{\overline{%
\zeta }}\overline{\left( X_{a}^{i}\right) }\right) $, we also have
\begin{equation}
\int_{\partial M_{+}}d\overline{\zeta }\left( \sum_{a=1}^{N}\frac{B}{2}%
\overline{\left( Y_{ia}\right) }\overline{\left( F_{a}^{i}\right) }\right)
\qquad \longleftrightarrow \qquad \int_{SM_{+}}d^{2}\upsilon d^{2}\overline{%
\mathrm{\theta }}\text{ }\overline{\mathcal{W}}\left( \Phi _{i}^{+}\right) .
\end{equation}%
Now, considering two chiral superfields $\Phi _{1}=\Phi _{1}\left( \upsilon
_{\pm },{\small \theta }_{\pm \text{{\small 1/2}}}\right) $ and $\Phi
_{2}=\Phi _{2}\left( \upsilon _{\pm },{\small \theta }_{\pm \text{{\small 1/2%
}}}\right) $ with ${\small \theta }$- expansions,
\begin{eqnarray}
\Phi _{1} &=&Y_{1}+{\small \theta }_{\text{{\small +1/2}}}{\small \psi }_{%
\text{{\small -1/2}}}+{\small \theta }_{\text{{\small -1/2}}}{\small \psi }_{%
\text{{\small +1/2}}}+{\small \theta }_{\text{{\small +1/2}}}{\small \theta }%
_{\text{{\small -1/2}}}F_{1},  \notag \\
\Phi _{2} &=&Y_{2}+{\small \theta }_{\text{{\small +1/2}}}{\small \varphi }_{%
\text{{\small -1/2}}}+{\small \theta }_{\text{{\small -1/2}}}{\small \varphi
}_{\text{{\small +1/2}}}-{\small \theta }_{\text{{\small +1/2}}}{\small %
\theta }_{\text{{\small -1/2}}}F_{2},
\end{eqnarray}%
with $Y_{i}$ and $F_{i}$ being the bosonic complex fields, we can build the
superpotential associated with the boundary QFT$_{2}$. We have,%
\begin{equation}
\int_{SM_{-}}d^{2}\mathrm{\theta }\left( \sum_{a=1}^{N}\frac{B}{2}\Phi
_{a1}\Phi _{a2}\right) =-\sum_{a=1}^{N}\left( \frac{B}{2}%
Y_{a1}F_{a2}-Y_{a2}F_{a1}\right) ,
\end{equation}%
which can be also written a covariant form as $\frac{B}{2}%
\sum_{a=1}^{N}\left( Y_{ia}F_{a}^{i}\right) $.

\qquad In the end of this section, we want to note that it would be
interesting to push further the similarity between the fluid of D-strings
and the usual FQH systems. As a next step, it is important to build the
ground state $|\Phi _{0}>$ of the quantized D-string model which may be done
by extending the matrix model approach of Susskind and Polychronakos. Recall
in passing that the fundamental wave function of standard FQH system on
plane with filling fraction $\nu =\frac{1}{k}$ is described by the Laughlin
wave,%
\begin{equation}
\Phi _{L}\left( x_{1},...,x_{N}\right) \sim \dprod\limits_{a<b=1}^{N}\left(
x_{a}-x_{b}\right) ^{k}e^{-B\sum_{a=1}^{N}\left\vert x_{a}\right\vert ^{2}}.
\end{equation}%
This wave function, which has been conjectured long time ago by Laughlin has
been recently rederived rigorously in \cite{16} by using matrix model
method. Notice that under permutation of particles, the wave function
behaves as,%
\begin{equation}
\Phi _{L}\left( x_{1},..,x_{a},..,x_{b},.,x_{N}\right) =\left( -\right)
^{k}\Phi _{L}\left( x_{1},..,x_{b},..,x_{a},.,x_{N}\right) .
\end{equation}%
Symmetry property of this function requires that $k$ should be a positive
odd integer for a system of fermions and an even integer for bosons.

\section{Conclusion and outlook}

\qquad In this paper, we have developed a gauge field theoretical model
proposal for a classical fluid of D-strings running in conifold and made
comments on its quantum behaviour. The field action $\mathcal{S}_{DSF}$ of
this classical conifold model, in presence of a strong and constant RR
background magnetic field \textrm{B}, exhibits a set of remarkable features.
It is a complex holomorphic functional $\mathcal{S}_{DSF}\left[ \mathcal{X},%
\mathcal{Y},\mathcal{C}_{0},\mathcal{\Lambda }\right] =\int_{T^{\ast
}S^{3}}d^{3}v$ $\mathcal{L}_{DSF}\left( \mathcal{X},\mathcal{Y},\mathcal{C}%
_{0},\mathcal{\Lambda }\right) $ with $\mathcal{L}_{DSF}=\mathcal{L}%
_{DSF}\left( \mathcal{X},\mathcal{Y},\mathcal{C}_{0},\mathcal{\Lambda }%
\right) $ given by,%
\begin{eqnarray}
\mathcal{L}_{DSF} &=&i\frac{\mathrm{B}_{RR}}{2\mu }\left( \mathcal{Y}%
_{i}\partial _{0}\mathcal{X}^{i}-\mathcal{X}^{i}\partial _{0}\mathcal{Y}%
_{i}\right) +\frac{\mathrm{B}_{RR}}{2\mu }\mathcal{\Lambda }\left( \mathcal{Y%
}_{i}\mathcal{X}^{i}-\mu \right)  \notag \\
&&-\frac{\mathrm{B}_{RR}}{\mu }\mathcal{C}_{0}\left( D_{++}\mathcal{Y}%
_{i}D_{--}\mathcal{X}^{i}-D_{--}\mathcal{Y}_{i}D_{++}\mathcal{X}^{i}\right) ,
\label{c1}
\end{eqnarray}%
where $\mathcal{\Lambda }$ is a Lagrange gauge field capturing the conifold
hypersurface. By setting $\left\{ F,G\right\} _{+-}=\left( D_{++}F\right)
\left( D_{--}G\right) -\left( D_{--}F\right) \left( D_{++}G\right) $ and
using general properties of the Poisson bracket, in particular antisymmetry
and Jacobi identity as well as the property,%
\begin{equation}
\mathcal{C}_{0}\left\{ \mathcal{Y}_{i},\mathcal{X}^{i}\right\} _{+-}=-%
\mathcal{Y}_{i}\left\{ \mathcal{C}_{0},\mathcal{X}^{i}\right\} _{+-}-%
\mathcal{C}_{0}\mathcal{Y}_{i}\mathcal{X}^{i}+\left(
D_{++}J_{--}+D_{--}J_{++}\right) ,
\end{equation}%
with $J_{\pm \pm }=\pm \left( \mathcal{C}_{0}\mathcal{Y}_{i}D_{\pm \pm }%
\mathcal{X}^{i}\right) $, the above holomorphic Lagrangian density $\mathcal{%
L}_{DSF}$ can be also put into a gauge covariant way as follows,%
\begin{equation}
\mathcal{L}_{DSF}=i\frac{\mathrm{B}_{RR}}{2\mu }\left( \mathcal{Y}_{i}%
\mathcal{D}_{0}\mathcal{X}^{i}-\mathcal{X}^{i}\mathcal{D}_{0}\mathcal{Y}%
_{i}\right) +\frac{\mathrm{B}_{RR}}{2\mu }\Lambda \left( \mathcal{Y}_{i}%
\mathcal{X}^{i}-\mu \right) ,
\end{equation}%
with $\mathcal{D}_{0}\mathcal{X}^{i}=\partial _{0}\mathcal{X}^{i}+i\left\{
\mathcal{C}_{0},\mathcal{X}^{i}\right\} _{+-}$. The presence of the Poisson
bracket $\left\{ \mathcal{C}_{0},\ast \right\} _{+-}$ in the gauge covariant
derivative $\mathcal{D}_{0}$ is a signal of non commutative gauge theory in
the same spirit as in Susskind description of Laughlin fluid. The basic
difference is that instead of a $U\left( 1\right) $ gauge group, we have
here a holomorphic $\mathbb{C}^{\ast }$ gauge symmetry acting on scalar
field as $\delta \mathcal{\Phi }=\left\{ \mathcal{\lambda },\mathcal{\Phi }%
\right\} _{+-}$ and $\delta \mathcal{C}_{0}=\partial _{0}\mathcal{\lambda }%
+i\left\{ \mathcal{C}_{0},\mathcal{\lambda }\right\} _{+-}$ with $\mathcal{%
\lambda }$ being the gauge parameter. Moreover, thinking about the D-string
field variables as
\begin{eqnarray}
\mathcal{X}^{i} &=&x^{i}+\mu \mathcal{C}_{+}^{i},  \notag \\
\mathcal{Y}_{i} &=&y_{i}-\mu \mathcal{C}_{-i},
\end{eqnarray}%
where the gauge fields $\mathcal{C}_{\pm }^{i}$\ describe fluctuations
around the static solution, $\mathcal{L}_{DSF}$ can be put in the form (\ref%
{hcs}) defining a complex holomorphic extension of the usual non commutative
Chern-Simons gauge theory. Notice that the role of the non commutative
parameter $\theta $ of usual FQH liquid is now played by the complex modulus
$\mu $ of the conifold in agreement with the observation of $\cite{9}$. The
topological gauge theory derived in this paper may be then thought of as
enveloping Susskind description of fractional quantum Hall fluid in Laughlin
state. The latter follows by restricting the conifold analysis to its
Lagrangian sub-manifolds by using eqs(\ref{02}). From this view the
D-strings fluid constitues a unified description of systems of FQH fluids in
real three dimensions, in particular those involving $R\times S^{2}$ and $%
S^{3}$\ geometries recovered as real slices of the conifold. The first
geometry is obtained by restricting world sheet variable $\xi =t+i\sigma $
to its real part and the second geometry is recovered by identifying $\xi $
with $\sigma $; that is a periodic time. For instance, the restriction of eq(%
\ref{c1}) to the real three sphere reads as
\begin{eqnarray}
\mathcal{L}_{FQH}^{\func{real}} &=&\frac{\func{Re}\left( \mathrm{B}%
_{RR}\right) }{2\func{Re}\left( \mu \right) }\left[ i\left( \overline{%
\mathcal{X}}_{i}\partial _{0}\mathcal{X}^{i}-\mathcal{X}^{i}\partial _{0}%
\overline{\mathcal{X}}_{i}\right) -2\mathcal{C}_{0}D_{++}\overline{\mathcal{X%
}}_{i}D_{--}\mathcal{X}^{i}\right] \\
&&+\frac{\func{Re}\left( \mathrm{B}_{RR}\right) }{2\func{Re}\mu }\left[ 2%
\mathcal{C}_{0}D_{--}\overline{\mathcal{X}}_{i}D_{++}\mathcal{X}^{i}+%
\mathcal{\Lambda }\left( \overline{\mathcal{X}}_{i}\mathcal{X}^{i}-\func{Re}%
\mu \right) \right] ,  \notag
\end{eqnarray}%
where $\mathcal{X}^{i}=\mathcal{X}^{i}\left( \sigma ,x,\overline{x}\right) $%
, $\overline{\mathcal{X}}_{i}=\overline{\left( \mathcal{X}^{i}\right) }$, $%
\mathcal{C}_{0}=\overline{\mathcal{C}_{0}},$ $\mathcal{\Lambda }_{0}=%
\overline{\mathcal{\Lambda }_{0}}$ and
\begin{equation}
D_{++}=x^{i}\frac{\partial }{\partial \overline{x}^{i}},\qquad D_{--}=%
\overline{x}^{i}\frac{\partial }{\partial x^{i}},\qquad D_{0}=\left[
D_{++},D_{--}\right] .
\end{equation}%
This analysis may be also viewed as a link between, on one hand, topological
strings on conifold, and, on the other hand, non commutative Chern Simons
gauge theory as well as FQH systems in real three dimensions. It would be
interesting to deeper this relation which may be used to approach attractor
mechanism on flux compactification by borrowing FQH ideas. To that purpose,
one should first identify the matrix model regularization of the continuous
field theory developed in this paper. This may be done by extending the
results of $\cite{13,14}$ obtained in the framework of fractional quantum
Hall droplets. An attempt using matrix field variables valued in $GL\left( N,%
\mathbb{C}\right) $ representations is under study in $\cite{15}$, progress
in this direction will be reported elsewhere.

\begin{acknowledgement}
\qquad\ \newline
This research work is supported by the program Protars III D12/25, CNRST.
\end{acknowledgement}


\begin{thebibliography}{99}
\bibitem{1} L. Susskind, The Quantum Hall Fluid and Non-Commutative Chern
Simons Theory, hep-th/0101029

\bibitem{2} Simeon Hellerman, Leonard Susskind, Realizing the Quantum Hall
System in String Theory, hep-th/0107200

\bibitem{3} Dimitra Karabali, Electromagnetic interactions of higher
dimensional quantum Hall droplets, Nucl.Phys B726 (2005) 407-420,
hep-th/0507027

\bibitem{4} Aziz El Rhalami, El Hassan Saidi, NC Effective Gauge Model for
Multilayer FQH States hep-th/0208144, JHEP

\bibitem{5} A.El Rhalami, E.M. Sahraoui, E.H.Saidi, NC Branes and
Hierarchies in Quantum Hall Fluids, JHEP 0205 (2002) 004, hep-th/0108096

\bibitem{6} S.C. Zhang, Quantum Hall effect in higher dimensions, (Talk
given at the Conference on Higher Dimensional Quantum Hall Effect,
Chern-Simons Theory and Non-Commutative Geometry in Condensed Matter Physics
and Field Theory, 1-4/03/2005, AS-ICTP Trieste

\bibitem{7} Hirosi Ooguri, Andrew Strominger, Cumrun Vafa, Black Hole
Attractors and the Topological String, Phys.Rev. D70 (2004) 106007,
hep-th/0405146

\bibitem{8} Hirosi Ooguri, Cumrun Vafa, Erik Verlinde, Hartle-Hawking
Wave-Function for Flux Compactifications, hep-th/0502211

\bibitem{9} EL Hassan Saidi, Topological $SL\left( 2\right) $\ gauge theory
on conifold and non commutative geometry, Lab/UFR-HEP/0514, GNPHE/0514,
VACBT/0514

\bibitem{10} James Gates Jr, Ahmed Jellal, EL Hassan Saidi, Michael
Schreiber, Supersymmetric Embedding of the Quantum Hall Matrix Model, JHEP
0411 (2004) 075 hep-th/0410070

\bibitem{11} Kazuki Hasebe, Supersymmetric Quantum Hall Effect on Fuzzy
Supersphere, Phys.Rev.Lett. 94 (2005) 206802, hep-th/0411137

\bibitem{12} Sergei Gukov, Kirill Saraikin, Cumrun Vafa, A Stringy Wave
Function for an S\symbol{94}3 Cosmology, hep-th/0505204

\bibitem{13} Bogdan Morariu, Alexios P. Polychronakos, Fractional quantum
Hall effect on the two-sphere: a matrix model proposal, Phys. Rev. D 72:
125002, 2005, hep-th/0510034

\bibitem{14} EL Hassan Saidi, Topological matrix model proposal for Laughlin
wave and cousin state, Lab/UFR-HEP0517/GNPHE/0519/VACBT/0519

\bibitem{15} R. Ahl Laamara, L.B Drissi, E H Saidi, D-string fluid in
conifold: II. Matrix model for D-droplets, in preparation.

\bibitem{16} Simeon Hellerman, Mark Van Raamsdonk, Quantum Hall Physics =
Noncommutative Field Theory, JHEP 0110 (2001) 039, hep-th/0103179

\bibitem{17} E.H. Saidi, M. Zakkari, Superconformal geometry from the
Grassmann and Harmonic Analycities, Int.J.Mod.Phys.A6:3151-3173,1991 \&
Int.J.Mod.Phys.A6:3175-3200,1991.

\bibitem{18} Marcos Marino, Chern-Simons Theory and Topological Strings,
Rev.Mod.Phys. 77 (2005) 675-720, hep-th/0406005.
\end{thebibliography}
\end{document}